\documentstyle[preprint,aps,prb]{revtex}
\begin{document}
\draft

\title{
New Fundamental dHvA Frequency in Canonical Low-Dimensional
Fermi Liquids
}
\author{ A.S. Alexandrov$^1$ and A.M. Bratkovsky$^2$ }
\address{
$^{(1)}$ Department of Physics, Loughborough University, LE11 3TU, U.K.\\
$^{(2)}$Hewlett-Packard Laboratories, Basic Research, 3500~Deer~Creek 
Road,
Bldg. 26U, Palo Alto, California 94304-1392. }
\date{\today}
\maketitle

\begin{abstract}

We show that  a new fundamental period $P_{f}$ of dHvA
oscillations, which appears  along
with other ``forbidden'' combination frequencies in a
multi-band canonical Fermi-liquid,  is very robust with
respect to a finite smearing of Landau levels  and a  background of
non-quantized states. We  analyse the possibility of  measuring
small Fermi surface pockets with  the use of the ``forbidden'' frequencies. 

\end{abstract}
\pacs{71.25.Hc, 71.18.+y, 71.10.Ay, 71.70.Di, 71.10.Pm}

In our previous  Letter \cite{alebra} we predicted
qualitatively different   dHvA  oscillations
in a multiband $2D$ $bulk$ metal
with a fixed fermion density  [canonical ensemble ($CE$)] compared with
an open system where the  chemical potential  is fixed [grand
canonical ensemble ($GCE$)].
The effect appears because of  the
pinning of the chemical potential
to a fractionally occupied Landau level (LL). Since the chemical potential $\mu$
is {\em constant} across the entire sample, the number of electrons in
individual
bands would {\em oscillate} in antiphase if the total number of
electrons is kept fixed, as will the partial orbital moments.
Therefore, in a closed system there will be a
mechanism for different bands to talk to each other producing 
a dHvA signal with the fundamental frequency, $F=F_{1}+F_{2}+\cdots$, where
$F_{1,2,\cdots}$ are the ordinary dHvA frequencies of the individual bands.

Let us first briefly recall  the qualitative arguments which led us to  
the prediction of $P_{f}\equiv 1/F$.\cite{alebra}
In two dimensions the number of electrons which the LL can accommodate
...is  always equal to $pH$, where $p$ is a universal constant
[$p=e/(2\pi\hbar 
c)$].
Then the  dHvA resonance  condition for fixed $n_{e}$ is
 \begin{equation}
 {e(H+\Delta H)(N-1)\over{2\pi\hbar c}}=n_{e},
\end{equation}
where $n_e$ is the number of electrons per $cm^2$ and $N=1,2,3...$ is determined 
by
${eH N/{(2\pi\hbar c)}}=n_{e}.$
Combining these equations, we obtain the fundamental dHvA period
\begin{equation}
P_f\equiv {1\over F}={p\over n_e},
\label{eq:pfund}
\end{equation}
which is independent of the band structure. This is a period (in $1/H$) with
which a fractionally filled Landau level will be completely
occupied, independently of any band characteristics. 

A relative occupation number of the highest occupied LL is given by 
$ x = F/H-[F/H] $, 
where $[z]$ stands for the integer part of $z$.
Firstly, it is obvious that this quantity oscillates with frequency $F$.
Secondly, $x$ is relevant for only a {\em closed} system (Canonical Ensemble)
since there the chemical potential is pinned to the
highest partially occupied Landau level in the limiting
case of negligible damping. Therefore, in the Canonical Ensemble (CE)
the chemical potential
 oscillates with the same
fundamental frequency $F$ as the occupancy $x$. 
Since the crystal field is much larger than the 
magnetic energies involved, oscillations in the chemical potential will
cause a charge flow between different sheets of the Fermi surface in a
multiband metal. Thus, there will be a component of the 
orbital magnetic moment oscillating with the fundamental frequency $F$
{\em interfering} with the standard dHvA individual band frequencies.

This mechanism surely does not operate in a 
Grand Canonical Ensemble (GCE), where $\mu$ is fixed by an external reservoir.
Therefore, in the GCE we have just the standard dHvA frequencies.  As mentioned in
Ref.~[\cite{alebra}] the  pinning of a chemical
potential
is only possible in 2D. In a 3D metal the Landau levels spread out in  bands
and the effect disappears together with a pinning of the chemical potential.
In general, the  effect can be suppressed by any mechanism which
introduces a
finite density of states (DOS) on the top of a sequence of Landau levels.
Therefore the role  of the {\em background} density of states and of the
damping should be carefully analyzed. 

The background density of states (g$_{bkg}$) may correspond to
either a non-quantized band (e.g. an open orbit) or a
heavily damped band, or to the three-dimensional  corrections to
the energy spectrum, or  to a finite contact resistivity.
We shall also analyze a truly
multiband case (a system with 3 bands with wildly different individual
frequencies) which may be relevant for several current experiments. We
observe new combination  dHvA frequencies
appearing because of the chemical potential pinning.

Let us consider a multiband 2D electron gas with a number of bands
labeled by the index $\alpha=1,2,\dots$ In an external magnetic field $H$ 
the electronic states will be quantized with energies 
\begin{equation}
\epsilon_\alpha(n) = \Delta_\alpha + \hbar\omega_\alpha(n + 
\frac{1}{2}),~~~n~=~0, 1, ...
\label{eq:LL}
\end{equation}
where  $\omega_\alpha=eB/(m_\alpha c)$ is the
cyclotron frequency, and $B=H+4 \pi M$ is the internal field. 
In the following we shall neglect the difference between $B$ and $H$
which is very small because of the small moment in real systems.\cite{martin}
Each level is degenerate, contains $pH$ states,
and is broadened by collisions with impurities into a Lorentzian with
Dingle width $\sim \hbar/\tau$.\cite{David84,Din51}

In the Canonical Ensemble ($n_e=const$)
the orbital moment is found from the Helmholtz free energy ${\cal F} =
{\cal F}(n_e,V,T)$ as 
\begin{equation}
M_{CE} = -\left( \partial {\cal F}\over \partial H\right)_{ T,V}=
-{k_BT}\int^\infty_{-\infty} d\epsilon
\left( {\partial g(\epsilon,H)\over \partial H}\right)_{T,V,\mu}
\ln\left( 1 + e^{ (\mu-\epsilon)/ k_BT } \right),
\label{eq:green}
\end{equation}
where $ g(\epsilon,H) = g_{L}(\epsilon,H) + g_{bkg}(\epsilon)$,
$g_{L}$ is the DOS 
derived from the Landau levels with account for collision 
broadening,\cite{YuB61} and 
$g_{bkg}(\epsilon)$ is the background density of states which may
arise because of the above-mentioned reasons.
We note that since the Gibbs free energy
$\cal F$ is related to the potential $\Omega$ as ${\cal F}=\Omega + \mu N$,
there is no explicit derivative $\partial \mu / \partial H$ in 
(\ref{eq:green}). In the Grand Canonical Ensemble we will find the moment
by the use of $\Omega(\mu,T,V)$ as 
$M_{GCE} = -\left( \partial \Omega/ \partial H\right)_{ T,V,\mu}$.
  The chemical potential $\mu$
is defined by the conservation of the total number of carriers, 
\begin{equation}
\int^{\infty}_{-\infty} d\epsilon (g_{L}(\epsilon) +
g_{bkg}(\epsilon)) = n_e
\label{eq:mu}
\end{equation}
To characterize the relative importance of different states, we shall
introduce an average density of states,
$\bar{g}^\alpha_{L} = pH/(\hbar \omega_h)=m^\alpha/(2\pi \hbar^2)$, 
which has the
dimension of $eV^{-1}cm^{-2}$.
We will then measure the background DOS with
respect to $\bar{g}_{L}$. Qualitatively, the behavior of $\mu$ is
quite clear. With the  increasing ratio $g_{bkg}/\bar{g}_{L}$ the
oscillatory component of the chemical potential vanishes as will
the fundamental frequency. 
It is important now to quantify the effect of the background DOS, 
which sets limits 
on experimental observation. It is also important to 
analyze a case with a few bands with largely different periods.

{}From now on we will measure all energies in the natural units of 
$\beta_0 F \equiv
2\pi\hbar^2n_e/m$, the moment per area in units of $\beta_0 n_e$, and the
thermodynamic potentials in units of $\beta_0 F n_e$,
where $\beta_0 = e\hbar/(mc)$.\cite{units} 
We shall  mark below all dimensionless quantities with a
tilde: $\tilde\mu=\mu/(\beta_0F)$, $ \tilde g_{bkg}=
g_{bkg}(\beta_0F/n_e)$, $\tilde{g}^\alpha_L=m_\alpha/m$.

Let us first consider the case of zero temperature in a clean limit
and a constant background DOS.  In this case 
$g_{L}(\epsilon)~=~pH\sum_\alpha \sum_{n=0}^\infty \delta(\epsilon - 
\epsilon_\alpha(n))$.
Then  in the  multiband case with the background DOS we will have the following
equation for $\mu$ in the CE:
\begin{equation}
\sum_{\alpha} N_\alpha +x + \tilde g_{bkg}\tilde\mu = {F\over H},
\label{eq:mu2}
\end{equation}
where $N_\alpha$ is the number of completely occupied LLs in a given band.
 We can write down the following expressions
for the thermodynamic potentials, up to an insignificant constant,
\begin{eqnarray}
{{\cal F} \over {\beta_0 F n_e}}  = {H\over F} \left( \sum^{occ}_{\alpha,n}
\tilde\epsilon_\alpha(n)  +  x\tilde\epsilon_\gamma(N_\gamma)\right) 
+ \frac{1}{2}\tilde g_{bkg}\tilde \mu^2,  \nonumber\\
{ \Omega \over \beta_0 F n_e}  = {H\over F} \left( \sum^{occ}_{\alpha,n}
(\tilde \epsilon_\alpha(n) -\tilde \mu) + x(\tilde 
\epsilon_\gamma(N_\gamma)-\tilde \mu) \right) 
- \frac{1}{2}\tilde g_{bkg}\tilde \mu^2,
\label{eq:zero}
\end{eqnarray}
where the sum runs over occupied LLs and $\gamma$ stands for the quantum
number of a fractionally occupied LL. We are working in the  semiclassical
limit, where $N_\alpha \gg 1$. In the CE the chemical potential will
oscillate about its mean value
\begin{equation}
\tilde \mu_{\rm mean} = \left(1+\sum_\alpha {m_\alpha \over m} \tilde 
\Delta_\alpha \right)
\left( {H\over F} \tilde g_{bkg} +\sum_\alpha {m_\alpha \over m} \right)^{-1},
\label{eq:mumean}
\end{equation}
where $m$ is the bare mass of the electron. The oscillation amplitude of the  
electron
density in individual bands will be 
$\Delta n_\alpha/n_\alpha = H/F_\alpha$ without background. A background
would tend to depin the chemical potential. For depinning of the chemical 
potential to occur,
the background density of states should actually become large, $g_{bkg} \gg 
g_{L}$,
so that the oscillations of the chemical potential will be  reduced as $\Delta
\mu
\propto H/(F\tilde g_{bkg})$. The following calculation substantiates this
result.

We compare in Fig.~1 the results of a numerical analysis of a two-band
model with mass ratio 1:4 and a band offset $\frac{1}{8} \beta_0F$.
In the top panel we present results for a system without background. 
The original dependence of a magnetic moment $M$ on the field $H$ is given
in the inset, and it looks like a superposition of two sawtooth curves
typical of the $M=M(F/H)$ dependence in a one-band case at $n_e=const$.
The fundamental period $F=H+L$ is clearly seen in the Fourier transform, as
well as its second and third harmonics ($2F$ and $3F$).\cite{jul}
Here $H$ and $L$ stands for the individual dHvA frequencies of a heavy
and a light bands, respectively.
 We also call 
attention to the presence of other combinational harmonics, though less
pronounced, like $H-L$, $F+L$, etc.

With a non-zero background density of states $g_{bkg}/g_{L}=5$,
the Fourier pattern becomes even better resolved with respect to the presence of 
a strong 
fundamental harmonic $F$, together with combination frequencies
$H-L$, $F+L$, and $2F-L$ harmonics.
We see that although the background is already very large $g_{bkg}/g_{L}
\gg 1$ compared to the light band, it is not strong enough to suppress the
fundamental frequency.
Only when $g_{bkg}/g_{L} \sim 25$ do we see that the intensity of
a fundamental harmonic reduces below the intensities of individual principal
harmonics $L$ and $H$. Then, with further increase of the background DOS,
the combinational frequencies die away and we recover a standard dHvA
behavior with two sets of individual harmonics (Fig.~1, bottom
panel). It demonstrates that the new fundamental frequency is very robust
with respect to the large  background density of states.

Another mechanism of reducing the fundamental harmonic would be
collision broadening. The effect is observable when
$\hbar/\tau \ll (\hbar \omega)_{min}$, which is a standard limitation for
the dHvA effect. Therefore, the new effect seems to be very robust and
does not require special conditions to be observed.

Now we are going to discuss a 3-band case relevant for current 
experiments.\cite{martin}  This is a  two-dimensional Fermi gas in
a quantum well
with three parabolic bands, where magnetic breakdown is completely
unlikely (trajectories are far from the Brillouin zone bondaries) 
and the Shoenberg effect (magnetic interaction) is negligible.\cite{martin}
In this example (Fig.~2) we have considered three bands with the ratio
of individual frequencies $F_A:F_B:F_C=5:13:53$ and equal masses. 
Again the fundamental harmonic $F=F_A+F_B+F_C$, marked $F=A+B+C$
in Fig.~2 (top panel), 
is the  second largest after $F_C$.
There is also a rich spectrum of combination frequencies.
One could have expected that the contribution of $F_A$ into the overall spectrum
to be always very small compared to the $B$ and $C$ bands, irrespective of
the ensemble.  On the contrary, the  band $A$,
although giving a small principal harmonic $F_A$  in CE,
produces a very strong combination frequency marked $C-A$ in
Fig.~2 (top panel) 
as it mixes with the band $C$ having largest weight. 

Incidentally,
this behavior opens up a potential possibility to detect low-frequency
modes analyzing ``forbidden'' dHvA harmonics like $C-A$ in our example.
It should be quite easy to detect such a mode by just looking at the 
positions of the
satellite peaks of the strongest harmonics. By attaching the system to a
reservoir (in the GCE) one eliminates all combination ``forbidden''
frequencies, as shown in Fig.~2 (bottom panel). 
Comparing CE and GCE one can see a
startling difference in behavior of the 3-band system in two regimes,
namely, the
presence of very extensive combinational harmonics in the Canonical
Ensemble (closed system) with the strong involvement of the low-frequency
harmonic, which nominally has a very small weight (Fig.~2, top panel).

Referring to other experiments we would like to point out that
there are a number of 2D systems, like ET salts, where ``forbidden''
frequencies may be observed. \cite{ET}
Thus, $\kappa$-(BEDT-TTF)$_2$Cu(NCS)$_2$ salt shows two 
orbits: $\alpha$, corresponding to a hole pocket, and $\beta$,
corresponding to simplest breakdown orbit. ``Forbidden'' frequencies
$F_\beta \pm F_\alpha$ appear on top of the standard ones
because of chemical potential pinning, as has been shown within our
model by M.~Nakano.\cite{nakano}

Considering  Sr$_{2}$RuO$_{4}$, which we discussed in our previous
paper, \cite{alebra}  we have to emphasize the possible relevance of our effect
 to this interesting $2D$ metal having the same crystal structure as
the high-$T_{c}$ cuprates. 
The band structure of   Sr$_2$RuO$_4$ remains an open issue. While
 the dHvA model \cite{mac} and standard
 band-structure calculations predict $two$ electron and $one$ hole pocket,
\cite{ogu}
 recent  photoemission spectroscopy results suggest  two $hole$ pockets
 \cite{yok} as in  our example.\cite{alebra} This difference as well
 as a significant mass enhancement is
 presumably due to a strong polaronic shift and the band narrowing  as
 we have discussed  for cuprates. \cite{alebra1,ale}

{}Finally, it is interesting to mention a possible effect of the finite
contact resistance $R$. Contacts are usually made to a sample to measure
Shubnikov-de Haas magnetoresistivity. Such a system is apparently
``open'' and should correspond to a Grand Canonical Ensemble. 
However, the sample is always capacitively coupled to a contact with 
some capacitance $C$.
If the ac frequency at which a measurement is carried out is higher
than $(RC)^{-1}$ then the exchange by electrons with a reservoir will
be suppressed. At these conditions the  system will behave in some aspects
as being
electrically insulated and combination ``forbidden'' frequencies
could be seen.

In conclusion, 
we have shown  that a new dHvA frequency, appearing in the Canonical
Ensemble of a two-dimensional electron gas, \cite{alebra}
is very robust with respect to the existence of smearing factors, 
like background density of states.
The background DOS
does not impose any stricter constraints on an
observation of the effect compared to the standard (GCE) case.
 We have demonstrated that the effect produces a tool for 
identifying small sheets of Fermi surface by analyzing ``forbidden''
dHvA frequencies. 
We are also pleased to see that our prediction of the new
fundamental dHvA frequency in the canonical multiband  $2D$ Fermi
liquid  has been confirmed in subsequent
 Fourier analyses\cite{jul,nakano} and the  relevant experiments are
currently underway. \cite{martin}

We highly appreciate  discussions with D. Shoenberg  on 
our Fourier analysis, with A. Mackenzie on Sr$_{2}$RuO$_{4}$, and we
thank M.Nakano for sending us his Fourier analysis. \cite{nakano}
We are grateful to M.Elliott \cite{martin} for information on his
recent experiments where combination frequencies have been detected.
     AMB acknowledges an interesting discussion of our previous Letter
\cite{alebra} with J. Singleton and his group in Oxford,
who has found afterwards  with co-workers\cite{ET-JS}
that a ``forbidden'' frequency 
$F_\beta - F_\alpha$ in $\kappa$-(ET)$_2$Cu(NCS)$_2$ salt
 can be explained by the mechanism that we have
suggested earlier.\cite{alebra}
The same conclusion has also  been independently reached by  
M. Nakano.\cite{nakano}

\figure{FIG.~1. dHvA oscillations in the canonical ensemble with two undamped
bands and background density of states $g_{bkg}$, which is measured w.r.t.
the averaged DOS $g_{L}$ of quantized Landau orbits 
(see text).
A strong signal comes from the fundamental frequency $F=L+H$
when $g_{bkg}/g_{L}=0$ (top panel). Relative intensity of the
$F$-harmonic
reduces with the increase of a background density of states 
($g_{bkg}/g_{L}=5$, middle panel),
but still remains very strong together with its higher harmonics.
These results are to be compared with the Grand Canonical Ensemble
(bottom panel)
where only standard harmonics are present.
 Units:
$m_H/m_L=4$, $\Delta_H - \Delta_L = \frac{1}{8} \beta_0 F$,
$g_L$ refers to ``light'' band.
Original orbital magnetization is shown in insets.
}

\figure{ FIG.~2. dHvA oscillations in 
the Canonical Ensemble with three undamped bands with 
frequencies $F_A:F_B:F_C = 5:13:53$ and equal masses.
A large number of combinational frequencies is seen in the Canonical
Ensemble (top panel). We note a strong mixing amplitudes  $C\pm A$,
much stronger than the principal low-frequency harmonic $A$. 
The fundamental harmonic
obeys a sum rule, $F = F_A + F_B + F_C$ and has the second strongest 
magnitude with very pronounced higher harmonics ($2F$) and ($3F$).
In the Grand Canonical Ensemble the spectrum consists of standard dHvA
individual harmonics (bottom panel).
Original orbital magnetization is shown in insets.
}


\noindent


\begin{references}

\bibitem{alebra}
A.S. Alexandrov and A.M. Bratkovsky, \prl
{\bf 76}, 1308 (1996); report mtrl-th/9509004 (1995).
\bibitem{David84} D. Shoenberg, {\em Magnetic Oscillations}
(Cambridge Univ. Press, Cambridge, 1984).
\bibitem{Din51} R.B. Dingle, Proc. Roy. Soc. {\bf A211}, 500,517 (1951).
\bibitem{YuB61} Yu. A. Bychkov, Sov. Phys. JETP {\bf 12}, 977 (1961).
\bibitem{martin} M. Elliott and R. Shepherd (private communication) 
have studied InGaAs quantum wells with three
parabolic bands with similar masses and found combination frequencies
in both dHvA and Shubnikov-de Haas measurements.
\bibitem{units} 
$F=n_e/p=41.4 T$ and $\beta_0 F = 4.87 meV$ for 
$n_e=10^{12} cm^{-2}$.

\bibitem{jul}
We are grateful to 
G.J. McMullan and D. Shoenberg, private communication (1996),
and M.Nakano, private communication (1996) who pointed out some
inaccuracies in our previous numerical Fourier analysis \cite{alebra} of
a Grand Canonical Ensemble (standard dHvA regime). Their analyses,
as well as a subsequent one by N. Harrison, J. Singleton, and F.Herlach
(unpublished, 1996) has confirmed our prediction of a new fundamental dHvA
frequency.
\bibitem{ET} F.A. Meyer {\em et al.} Europhys. Lett. {\bf 32}, 681 (1995).

\bibitem{nakano} M. Nakano, J. Phys. Soc. Jpn. (1996), to be published.
In this paper a fundamental harmonic $F_\alpha + F_\beta$ appears to
be the strongest one, whereas in our two-band case (Fig.~1) or three-band
case (Fig.~2) it is lower in energy than the strongest individual
harmonic. It should be checked that this is not due to 
numerical problems.

\bibitem{mac}
A.P. Mackenzie $et$ $al.$, \prl {\bf 76}, 3786 (1996).
\bibitem{ogu}
T. Oguchi, Phys. Rev. B {\bf 51}, 1385 (1995); D.J. Singh, \prb {\bf
52}, 1358 (1995).
\bibitem{yok}
T. Yokoy $et$ $al.$, \prl {\bf 76}, 3009 (1996).
\bibitem{alebra1}
A.S. Alexandrov, A.M. Bratkovsky, and N.F. Mott, \prl {\bf 72}, 1734
(1994); $ibid.$ {\bf 74}, 2840 (1995).
\bibitem{ale}
A.S. Alexandrov, \prb {\bf 53}, 2863 (1996).
\bibitem{ET-JS} 
N. Harrison, J. Caulfield, J. Singleton, P.H.P. Reinders,
F.Herlach, W. Hayes, M. Kurmoo, P. Day,
J. Phys. Condens. Matter {\bf 8},  5115 (1996). 
See also discussion by M.~Nakano.\cite{nakano}




\end{references}
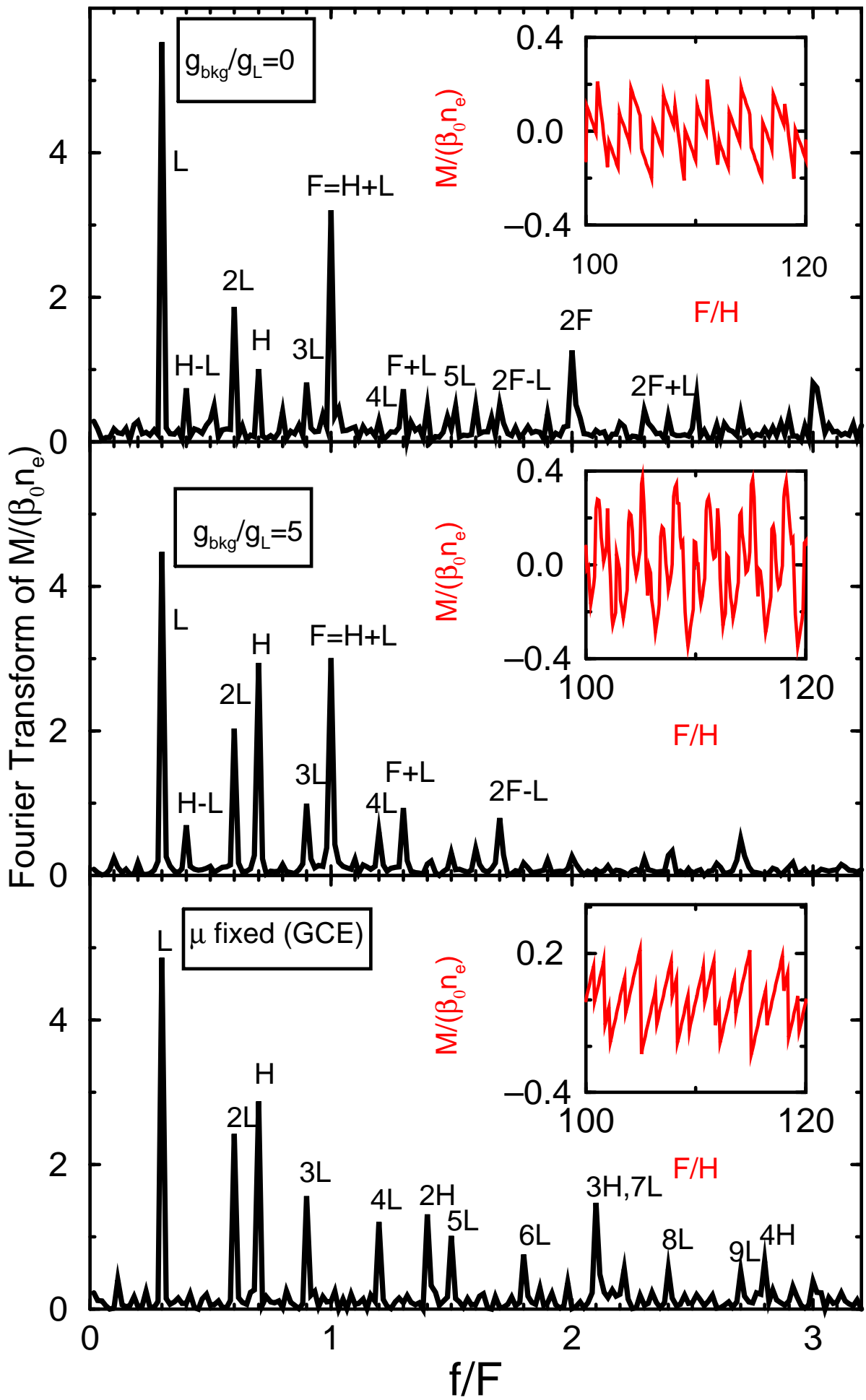
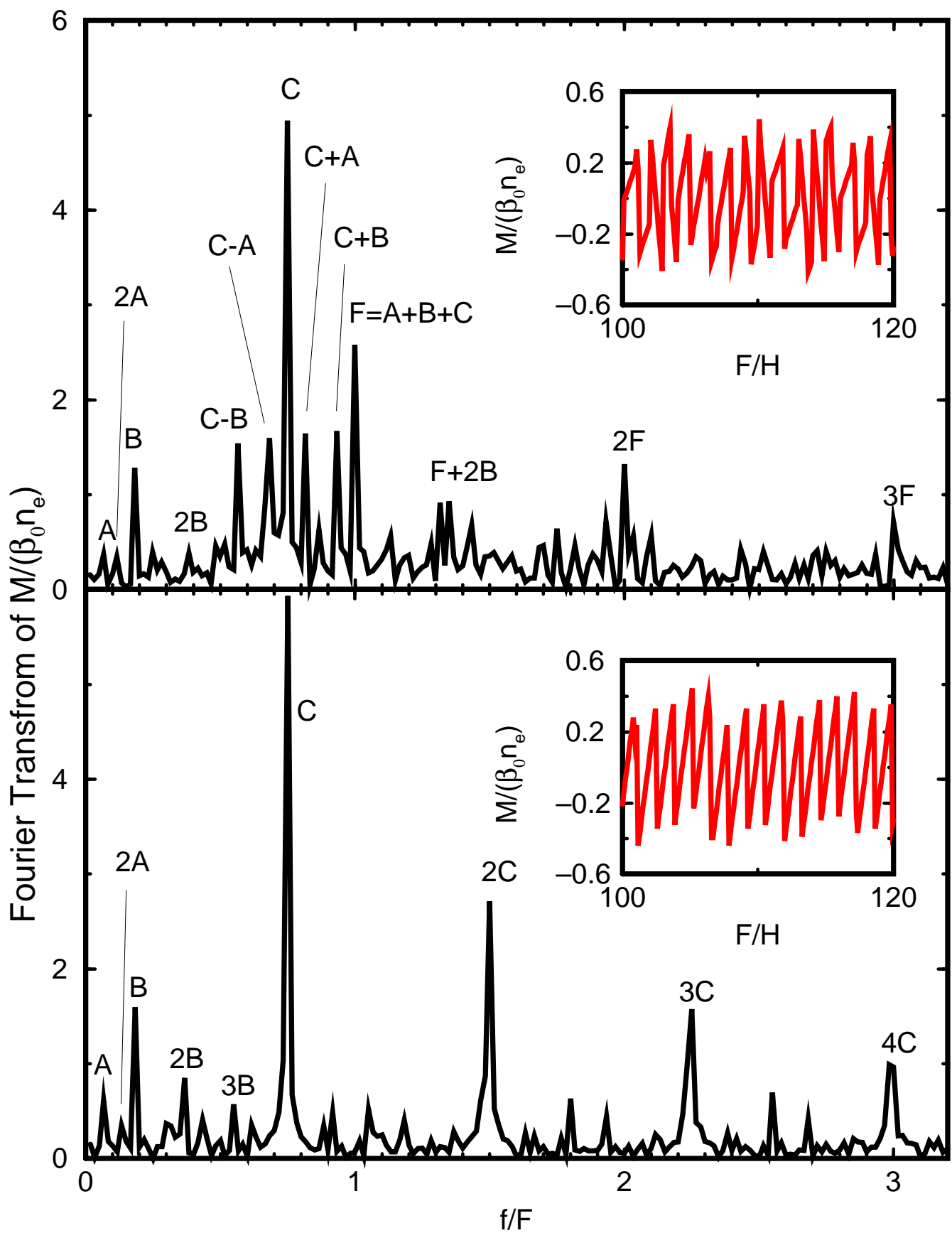
\end{document}